\documentclass[fleqn,usenatbib]{mnras}

\usepackage{newtxtext,newtxmath}

\usepackage[T1]{fontenc} 
\usepackage{amsmath} 

\DeclareRobustCommand{\VAN}[3]{#2}
\let\VANthebibliography\thebibliography
\def\thebibliography{\DeclareRobustCommand{\VAN}[3]{##3}\VANthebibliography}

\usepackage{graphicx}	
\usepackage{amsmath}	
\usepackage{orcidlink}

\newcommand{\edited}[1]{{#1}}

\newcommand{\AdrienRevision}[1]{{#1}}
\newcommand{\AdrienFinalRev}[1]{{#1}}

\title[Smuggling unnoticed]{Smuggling unnoticed: Towards a 2D view of water and dust delivery to the inner regions of protoplanetary discs}

\author[Houge et al.]{Adrien Houge$^{1, 2}$\thanks{E-mail: adrien.houge@sund.ku.dk}\orcidlink{0000-0001-8790-9011},
Sebastiaan Krijt$^{1}$\orcidlink{0000-0002-3291-6887},
Andrea Banzatti$^{3}$\orcidlink{0000-0003-4335-0900},
Geoffrey A. Blake$^{4}$\orcidlink{0000-0003-0787-1610},
Paola Pinilla$^{5}$\orcidlink{0000-0001-8764-1780},
\newauthor
Klaus M. Pontoppidan$^{6, 4}$\orcidlink{0000-0001-7552-1562},
Leon Trapman$^{7}$\orcidlink{0000-0002-8623-9703},
Joe Williams$^{1}$\orcidlink{0009-0008-8176-1974},
and Ke Zhang$^{7}$\orcidlink{0000-0002-0661-7517}
\\
$^{1}$Department of Physics and Astronomy, University of Exeter, Exeter, EX4 4QL, UK\\
$^{2}$Center for Star and Planet Formation, GLOBE Institute, University of Copenhagen, Øster Voldgade 5-7, DK-1350 Copenhagen, Denmark\\
$^{3}$Department of Physics, Texas State University, 749 North Comanche Street, San Marcos, TX 78666, USA\\
$^{4}$Division of Geological and Planetary Sciences, California Institute of Technology, MC 150-21, 1200 E California Boulevard, Pasadena, CA 91125, USA\\
$^{5}$Mullard Space Science Laboratory, University College London, Holmbury St Mary, Dorking, Surrey RH5 6NT, UK\\
$^{6}$Jet Propulsion Laboratory, California Institute of Technology, 4800 Oak Grove Drive, Pasadena, CA 91109, USA\\
$^{7}$Department of Astronomy, University of Wisconsin-Madison, Madison, WI 53706, USA
}

\date{Accepted 2025 January 8. Received 2025 January 8; in original form 2024 June 26}

\pubyear{\the\year{}}

\begin{document}
\label{firstpage}
\pagerange{\pageref{firstpage}--\pageref{lastpage}}
\maketitle

\begin{abstract}

Infrared spectroscopy, e.g., with JWST, provides a glimpse into the chemical inventory of the innermost region of protoplanetary discs, where terrestrial planets eventually form. The chemical make-up of regions inside snowlines is connected to the material drifting from the outer regions, which can be modeled with dust evolution models. However, infrared observations are limited by the high dust extinction in the inner disc, and only probes the abundances of gaseous species in the disc surface layers. As a result, the bulk mass of delivered volatiles is not directly relatable to what is measured through infrared spectra. In this paper, we investigate how the delivery of dust and ice after prolonged pebble drift affects the observable reservoir of water vapor in the inner disc. We develop a 1+1D approach based on dust evolution models to determine the delivery and distribution of vapor compared to the height of the $\tau = 1$ surface in the dust continuum. We find that the observable column density of water vapor at wavelengths probed by JWST spans many orders of magnitude over time, exhibiting different radial profiles depending on dust properties, drift rate, and \AdrienRevision{local processing}. In the presence of a traffic-jam effect \AdrienRevision{inside} the snowline, the observable vapor reservoir \AdrienRevision{appears constant} in time despite the ongoing delivery by pebble drift\AdrienFinalRev{, such that water is effectively smuggled unnoticed}. Differences in measured column densities then originate not only from variations in \AdrienRevision{bulk vapor content,} but also from differences in the properties and distribution of dust particles.

\end{abstract}

\begin{keywords}
protoplanetary discs -- planets and satellites: composition -- planets and satellites: formation -- methods: numerical 
\end{keywords}

\section{Introduction} 
\label{sec:intro}

The inner region of protoplanetary discs ($r < 5 \mathrm{~au}$) is the birthplace of terrestrial planets, and hold key clues to understand how Earth-like planets form, and what may shape their habitability. These regions are, however, {difficult to spatially resolve}, even with the high angular resolution of the Atacama Large Millimeter/submillimeter Array (ALMA). To probe the chemical inventory of the inner few astronomical units of discs, we thus generally use infrared observations, where the high temperature at these locations allow for the measurements of rich spectra featuring numerous \AdrienRevision{rotational and} ro-vibrational transitions of small carbon and oxygen-bearing molecules \citep{carr2008organics, carr2011organics, salyk2008h2o, salyk2011spitzer, salyk2019high, pontoppidan2010spitzer, mandell2012first, walsh2015molecular, banzatti2017depletion, banzatti2023kinematics}. A main complication at these wavelengths is dust extinction, which obscures the midplane and limits line emission to the upper layers of the disc atmosphere.

Infrared {spectra} are typically fitted either with slab models, estimating the temperature, column density and emitting area of given molecules \citep[e.g.,][]{carr2008organics}, or complex thermo-chemical models, such as \texttt{ProDiMo} \citep[e.g.,][]{woitke2009radiation} and \texttt{DALI} \citep[e.g.,][]{bruderer2012warm, bruderer2013survival}, based on otherwise static dust and gas distributions, and elemental abundances. Early results with the Spitzer Space Telescope taught us that the infrared spectra of T Tauri stars exhibit a significant diversity, varying with disc and star properties \citep[e.g., review by][]{pontoppidan2014volatiles}, with for example the water/HCN ratio (a proxy for the C/O ratio) varying significantly with stellar mass \citep{pascucci2009different, pascucci2013atomic}. Combining these infrared spectra with sub-mm interferometric data, which allows to estimate global disc properties, population-level trends were also found, showing that inner disc line flux ratios {(e.g., water and HCN)} correlate with dust disc mass \citep{najita2013hcn} and/or mm-dust disc size \citep{banzatti2020hints}.

More recently, the increased sensitivity and spectral resolution of the James Webb Space Telescope (JWST) has allowed us to dive deeper than ever into the chemical reservoir of the inner disc, with early results confirming the \edited{diversity in} line flux ratios \citep[e.g.,][]{perotti2023water, tabone2023rich, xie2023water, grant2023minds, tannus2023xue, grant2024minds}. \edited{From the analysis of four discs observed with MIRI (two compact, two extended with multiple dust gaps), \citet{banzatti2023jwst} found that their water spectra are visibly different in the excitation of low-energy lines, which are stronger in the compact discs and show excess emission that is reproduced by a cool water reservoir \AdrienRevision{($190-400 \mathrm{~K}$)} that is instead reduced in the extended discs}. They hypothesised that the cold water reservoir originates from the drift and subsequent sublimation of icy pebbles from the outer disc \citep{cuzzi2004material}. In this scenario, discs without substructures would be associated with a higher flux of icy pebbles, also leading to a decrease of the dust disc size with ongoing drift \citep[e.g.,][]{pinilla2012trapping, appelgren2020dust, toci2021secular, appelgren2023disc}, and a greater delivery of water vapor in the innermost regions. The presence of a more massive cold water vapor reservoir seems to be confirmed in the light of new slab models results on a sample of four compact and three extended discs \AdrienRevision{\citep[][]{romero2024retrieval}}.

Disc-wide models designed to study the interplay between dust evolution (e.g., coagulation, drift) and disc chemistry (e.g, elemental and molecular abundances, abundance ratios) confirm that the chemical make-up of regions inside major sublimation fronts (mainly the $\mathrm{H_2O}$ and $\mathrm{CO}$ snowlines) can be affected by prolonged pebble drift \citep{booth2017chemical, krijt2018transport, booth2019planet, krijt2020co, kalyaan2021linking, kalyaan2023effect}. This is opening a new paradigm in the field, where the properties of terrestrial planets do not only rely on local processes, but are deeply connected to what is happening in the outer regions where drifting icy-particles originate. So far, however, most these models only consider the radial evolution of dust particles and vapor, and it remains challenging to connect the drift and evaporation of pebbles (happening in the midplane) to the column density that may be measured with infrared line fluxes coming from the disc atmosphere. In particular, previous modeling efforts are focused on how much total water mass can be delivered to the inner disc \citep{kalyaan2021linking, kalyaan2023effect}, but neglect the impact of all the dust that will also be delivered along with pebble drift, which may accumulate inside the snowline and increase the optical depth and dust extinction effects \citep{bosman2023potential}. 

In this paper, we seek to study the co-evolution of dust and ice, and their delivery in the inner disc. We investigate how considering both enrichment in dust and ice may impact the water vapor reservoir that is ultimately observed. We will perform dust coagulation and evolution simulations with \AdrienRevision{\texttt{chemcomp} \citep{schneider2021how}}, to compute the dust size distribution and radial pebble flux as a function of time and distance to the star. Then, using a 1+1D approach, we will determine the dust distribution in the vertical dimension, from which we will derive the evolution of the $\tau = 1$ surface with ongoing pebble drift. Taken together, we will estimate what fraction of the evolving water vapor reservoir may be observable in the inner disc above the $\tau = 1$ surface. Given the sensitivity of pebble drift and inner disc dust density on the fragmentation limit $v_\mathrm{frag}$, but the ongoing debates concerning its value for dry silicate and ice-rich particles \citep[e.g.,][]{gundlach2014stickiness, gundlach2018tensile}, we will consider three dust models with low, high, or composition-dependent fragmentation velocity {\citep[similar to e.g.,][]{banzatti2015direct}.}

\AdrienRevision{This paper is organized as follows. In Section \ref{sec:methods}, we detail our dust and vapor evolution model based on \texttt{chemcomp}, along with our 1+1D treatment to compute the dust optical depth, observable vapor reservoir, and local UV and chemical processing. We present our results in Section \ref{sec:results}, discuss our findings in Section \ref{sec:discussion_conclusion}, before giving our conclusions in Section \ref{sec:conclusion}.}

\section{Methods} 
\label{sec:methods}

\begin{figure*}
\begin{center}

\includegraphics[width=\textwidth]{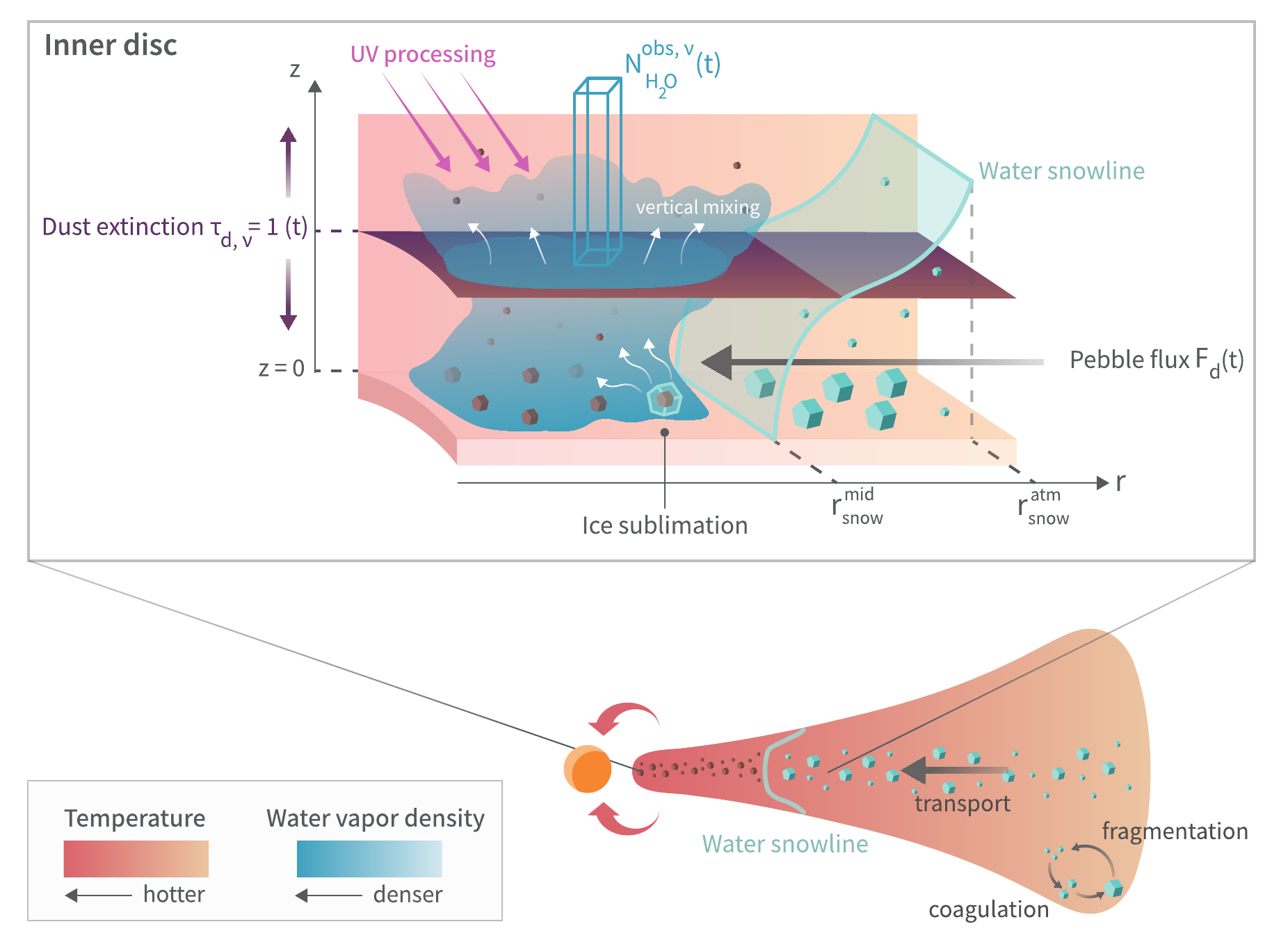}
    \caption{Schematic representing a protoplanetary disc and a zoom on the inner regions close to the water snowline. Outside the snowline, water ice is deposited on the surface of dust particles. The coagulation of icy grains leads to the formation of larger icy pebbles that are well-settled in the midplane and drift inwards efficiently, resulting in a time-dependent pebble flux $\mathcal{F}_\mathrm{d}$. When crossing the snowline, water sublimates and diffuses, enriching the inner disc with water vapor. \AdrienRevision{Vertical mixing brings water vapor in the disc surface layers, where it may be processed by UV irradiation and chemical reactions.} The refractory dust leftovers contribute to maintaining a high dust optical depth in the inner disc. Due to the high dust densities and extinction, only a fraction of the inner disc vapor reservoir may be observable, corresponding to the material above the $\tau_\nu = 1$ surface.}
    \label{fig:schematic_innerdisc}
\end{center}
\end{figure*}

In this section, we describe our approach to simulate dust coagulation and transport, estimate the vertical distribution of dust and vapor, and determine the observable column density of water vapor. Our approach is illustrated in Fig.~\ref{fig:schematic_innerdisc}.

\subsection{Evolution of the bulk dust and water content}

\AdrienRevision{To study the delivery of dust and water ice to the inner disc, we use the one-dimensional code \texttt{chemcomp} \citep{schneider2021how}, designed to simulate the coagulation and radial transport of dust \citep[with the two-pop-py algorithm, see][]{birnstiel2012simple}, along with the transport (advection, diffusion), sublimation, and condensation of volatile species. For this work, we adapt the code to include the evolution of only two species: water and refractory dust. Outside of the water snowline (i.e., for $T < 150 \mathrm{~K}$), water is deposited on the surface of refractory grains, such that their dynamical evolution is coupled. We fix the initial water-to-refractory mass fraction to $\delta_\mathrm{w2r, 0}=0.2$, which is comparable with measurements of comet 67P/Churyumov-Gerasimenko \citep{patzold2016homogeneous}. We simulate the disc with standard \texttt{chemcomp} settings, over a discretised grid of $500$ radial cells log-spaced between $0.1\mathrm{~au}$ and $1000~\mathrm{au}$. The initial gas surface density profile follows a tapered power-law \citep[]{lynden1974evolution}} set by the scaling radius $r_\mathrm{0}=100 \mathrm{~au}$ and initial disc mass $M_\mathrm{disc,0} = 0.05 \mathrm{~M_\odot}$ \citep[see more details about standard \texttt{chemcomp} settings in][]{schneider2021how}. \edited{The stellar mass is set to $M_* = 1 \mathrm{~M_\odot}$}. We parameterize the disc turbulent viscosity following \citet{shakura1973black}, setting the turbulent parameter to $\alpha_\mathrm{T} = 10^{-3}$ \citep{rosotti2023empirical}. \edited{The radial diffusion $\delta_\mathrm{rad}$, vertical stirring $\delta_\mathrm{vert}$, and turbulent mixing $\delta_\mathrm{turb}$ (which controls collision velocities between dust particles), are all set to the value of the turbulent parameter $\alpha_\mathrm{T}$ \citep[see e.g.,][for a study where these quantities differ]{pinilla2021growing}}. We assume the disc to be passively irradiated, such that the dust and gas temperature in the midplane follows
\begin{equation}
\label{eq:midplane_temperature}
T_\mathrm{mid} (r) = \left(\frac{\phi L_\mathrm{*}}{8 \pi\sigma_\mathrm{SB} r^2}\right)^{1/4},
\end{equation}
where $\phi=0.05$ is the disc opening angle and \AdrienRevision{$L_\mathrm{*}=4~\mathrm{L_\odot}$ is the stellar luminosity. With this temperature profile, the midplane water snowline is located} at $r_\mathrm{snow}^\mathrm{mid} \approx 2.2 \mathrm{~au}$. At such small distances from the central star, the temperature structure may deviate from the passively irradiated case, e.g., due to self-shadowing effects or viscous heating. Viscous heating, if significant, leads to an increase in the midplane temperature of the inner regions, pushing the midplane snowline further out and resulting in a more complex 2D temperature profile \citep[e.g.,][]{min2011thermal}. To properly investigate the impact of these effects on the observable water column, the time evolution of the discs's mass accretion rate, dust content and opacity will have to be solved self-consistently. This will be the focus of future work. 

\AdrienRevision{To compute the evolution of the dust size distribution, \texttt{chemcomp} \citep[see details in][]{schneider2021how} uses the two-pop-py algorithm \citep{birnstiel2012simple}, that reproduces well the results from detailed coagulation models that include dust coagulation, fragmentation, and radial drift \citep[e.g., DustPy, ][]{stammler2022dustpy}, with much lower computational expenses. The initial population of dust grains consists in $0.1 \mathrm{~\mu m}$ monomers well coupled to the gas. Dust particles coagulate with time, depending on the growth timescale that is itself dependent on the local dust density, up to a maximum size $a_\mathrm{max}(r, t)$ either set by the fragmentation barrier or the radial drift barrier. The former is controlled by the fragmentation velocity $v_\mathrm{frag}$, above which collisions between particles outcome on fragmentation rather than growth.} We adopt three models concerning the \AdrienRevision{value of the} fragmentation velocity of dust particles: (i) the \textit{resistant} dust model, where $v_\mathrm{frag} = 10 \mathrm{~m~s^{-1}}$ throughout the disc, (ii) the \textit{composition-dependent} model, where ice-covered particles benefit from a higher resistance to fragmentation than dry silicates, such that $v_\mathrm{frag} = 10 \mathrm{~m~s^{-1}}$ outside the snowline and $v_\mathrm{frag} = 1 \mathrm{~m~s^{-1}}$ inside the snowline, in agreement with previous findings \citep{dominik1997physics, supulver1997sticking, gundlach2014stickiness}, and (iii) the \textit{fragile} dust model, where $v_\mathrm{frag} = 1 \mathrm{~m~s^{-1}}$ throughout the disc, in agreement with more recent laboratory experiments on icy aggregates \citep{gundlach2018tensile, musiolik2019contacts}. \edited{Multi-wavelength observations of nearby discs also appear to favor fragile grains \citep{jiang2024grain}, although the exact constraints on $v_\mathrm{frag}$ depend on the assumed turbulence levels and grain porosity}. 

\AdrienRevision{When inward-drifting icy pebbles cross the snowline, they release water vapor and refractory grains. We assume refractory grains to remain intact despite the sublimation of their ice content, in light of recent analysis of the outbursting disc V883 Ori \citep{houge2024surviving}. Water vapor is then evolved by \texttt{chemcomp} as a trace species that advects and diffuses within the bulk gas. It may move inwards, leading to mass loss as it accretes onto the central star, or diffuse outwards across the snowline, where it re-condenses on dust particles.}

\subsection{Evolution of the observable water reservoir}
\label{sec:methods_2D}

\AdrienRevision{The \texttt{chemcomp} models return radial profiles of the bulk gas $\Sigma_\mathrm{g}(r,t)$, water $\Sigma_\mathrm{H_2O}(r,t)$, and dust $\Sigma_\mathrm{d}(r,t)$ surface density as a function of time, as well as the maximum grain size $a_\mathrm{max}(r,t)$ and solid composition. Here we describe how we turn these outputs into estimates of the amount of \emph{observable} water vapor in the disc surface.} 

\subsubsection{Radial and vertical dust distribution}
\label{sec:vertical_distrib_methods}

We employ a 1+1D approach where we construct a vertical grid at every radial bin $r$, \AdrienRevision{containing $64$ cells} going from the midplane $z/r = 0$ to $z/r = 0.8$, the upper layer of the disc atmosphere. \AdrienRevision{We can derive the radial and vertical dust density distribution from the \texttt{chemcomp} outputs by: (1) assuming that at a given radius $r$ the vertically-integrated dust size distribution is represented by an MRN-like power-law $n(a) \propto a^{-q}$ with $q=3.5$ \citep[][]{mathis1977size} between a minimum size of $0.1 \mathrm{~\mu m}$ and a maximum size $a_\mathrm{max}(r,t)$, and (2) for every individual grain size the vertical distribution follows an equilibrium between settling and turbulent mixing.} In that case, for each grain size characterized by a (midplane) Stokes number $\mathrm{St}$ \citep[e.g.,][]{pinilla2021growing},
\begin{equation}
\rho_{\mathrm{d}}(r, z, \mathrm{St})= \frac{\Sigma_\mathrm{d}(r, \mathrm{St})}{\sqrt{2 \pi} h_{\mathrm{d}}(r, \mathrm{St})} \exp \left\{-\frac{z^{2}}{2 h_{\mathrm{d}}^{2}(r, \mathrm{St})}\right\},
\label{eq:dust_density}
\end{equation}
where $h_\mathrm{d}=h_\mathrm{g} \sqrt{\delta_\mathrm{vert}/(\delta_\mathrm{vert}+\mathrm{St})}$ is the dust scale-height\footnote{\AdrienRevision{We performed tests with an updated dust scale-height prescription that includes stronger settling in the surface regions where St increases \citep[see][for details]{birnstiel2016dust, woitke20232d} and found that while the dust densities above $z/R \sim 0.2$ were impacted, there was no noticeable change in the height of the $\tau_\mathrm{20\mu m}=1$ layer for all simulations considered here.}} \citep{dubrulle1995dust}, $\Sigma_\mathrm{d}$ is the dust surface density, $h_\mathrm{g} = c_\mathrm{s}/\Omega$ is the gas pressure scale-height, $c_\mathrm{s} = \sqrt{k_{\mathrm{B}} T / m_{\mathrm{g}}}$ is the sound-speed, $\Omega=\sqrt{G M_{*} / r^{3}}$ is the Keplerian frequency, $k_\mathrm{B}$ is the Boltzmann constant, $m_{\mathrm{g}} = 2.34$ amu is the mean molecular mass, and $T$ is the midplane temperature (Equation~\ref{eq:midplane_temperature}). While using Equation~\ref{eq:dust_density} to distribute different grain sizes vertically results in a dust size distribution that varies with $z$, the dust size distribution \AdrienRevision{built with \texttt{chemcomp} assumes} midplane values for the relative velocities between particles, and for the fragmentation barrier. As such, the results of a full 2D model could differ from the approach taken here, as collision velocities would increase with $z$, potentially resulting in higher collision rates \citep[e.g.,][]{zsom2011outcome} and a shift in the fragmentation barrier \citep[e.g.,][figure 1]{krijt2016dust}, both of which could impact the local size distribution. For a recent 2D model that can capture these effects, see \citet{robinson2024cudisc}. Concerning the vertical structure of the gas density $\rho_\mathrm{g}(r,z)$, we also assume a vertical hydrostatic equilibrium and use Equation~\ref{eq:dust_density}, though replacing the dust scale-height $h_\mathrm{d}$ by the gas pressure scale-height $h_\mathrm{g}$.

\subsubsection{2D temperature profile and water snowline}
\label{sec:2D_temperature}
We employ a parametric 2D temperature profile for the gas temperature $T_\mathrm{g}$ as a function of stellar irradiation. Following \citet{dartois2003structure}, we express the gas temperature as
\begin{equation}
\label{eq:2D_gas_temperature}
T_\mathrm{g}(r, z)=\left\{\begin{array}{ll}
T_\mathrm{atm}+\left(T_\mathrm{mid}-T_\mathrm{atm}\right) \left[\cos\left(\dfrac{\pi z}{2 z_\mathrm{q}}\right)\right]^{2 \sigma} & \text { if } z<z_\mathrm{q} \\
T_{\mathrm{atm}} & \text { if } z \geqslant z_\mathrm{q}
\end{array} .\right.
\end{equation}
where $T_\mathrm{mid}$ is the midplane temperature determined by Equation~\ref{eq:midplane_temperature}, as in the dense midplane we can assume the dust and gas temperature to be equal, $T_\mathrm{atm} = 2 ~T_\mathrm{mid}$ is the atmosphere temperature, higher than the disc midplane notably due to heating by UV photons, $\sigma = 1$ is the stiffness parameter of the vertical profile, and $z_\mathrm{q}$ is the height at which the gas temperature reaches the atmospheric value. We fix $z_\mathrm{q} = 4 ~h_\mathrm{g}$, similarly to previous works \citep[e.g.,][]{dartois2003structure, williams2014parametric}. 

\AdrienRevision{With this temperature profile, water vapor still exists in the warmer disc surface layers outside the midplane snowline, such that vapor can be found up to $r\sim10\mathrm{~au}$ (see Fig.~\ref{fig:2D_dust_distribution}). Beyond this radius, even the disc atmosphere becomes too cold for water vapor to be present. Because \texttt{chemcomp} is a 1D code, however, all the water content outside the midplane snowline $r_\mathrm{snow}^\mathrm{mid}$ is treated as ice. When extending the \texttt{chemcomp} models to 2D, we thus convert the water ice density above the snowline ($T > 150 \mathrm{~K}$) into vapor. However, this may lead to a vertical gradient due to vertical settling, which should not affect gaseous species. We thus re-distribute the vapor density vertically to ensure $(\rho_\mathrm{H_2O}/\rho_\mathrm{g})(z_{T > 150 \mathrm{~K}}) = \mathrm{const.}$, while conserving the total water mass in the column.}

\subsubsection{Dust vertical optical depth}

The second key quantity of this paper is the dust optical depth, especially, the height at which it equals unity, as anything under this surface will be hidden from observations. We compute the dust optical depth at each point of the 2D grid with
\begin{equation}
\label{eq:optical_depth}
    \tau_\nu^\mathrm{d}(r, z, t) = \int_{z}^{\infty} \kappa_\nu^\mathrm{tot}(r, z', t) \rho_\mathrm{d}(r, z', t) dz,
\end{equation}
where $\kappa_\nu^\mathrm{tot}$ is the total opacity of the dust distribution (see Appendix \ref{sec:appendix_opacity}) and $\rho_\mathrm{d}$ is the dust density, obtained from \texttt{chemcomp} and expanded in the vertical dimension assuming dust particles are in a vertical equilibrium between settling and turbulent mixing (Equation~\ref{eq:dust_density}).

From \AdrienRevision{the bulk water vapor content present in the inner disc}, only a fraction will be observable, corresponding to the fraction of vapor above the height $z(\tau_\nu=1)$ at which the dust optical depth equals unity. We compute the observable column density $N_\mathrm{H_2O}^\mathrm{obs, \nu}$ \AdrienRevision{(in $\mathrm{cm^{-2}}$)} as
\begin{equation}
\label{eq:observable_column_density}
    N_\mathrm{H_2O}^\mathrm{obs, \nu} = \int_{z(\tau_\nu = 1)}^{\infty} 
    \dfrac{\rho_\mathrm{H_2O}}{m_\mathrm{H_2O}} ~dz,
\end{equation}
where $m_\mathrm{H_2O}$ is the mass of a water molecule. \edited{This approach neglects water emission that may originate from depths deeper than the $\tau = 1$ surface. We estimated that fraction by integrating the vapor density from the midplane, but adding an extinction factor $\exp{(-\tau_\nu)}$. We find that it agrees with the estimate given by Equation~\ref{eq:observable_column_density} within $\sim 10\%$.}

\subsubsection{Water abundances in the disc surface layers}
\label{sec:UV_loss_methods}

\AdrienRevision{We are primarily interested in the amount of water in the disc surface layers. While \texttt{chemcomp} returns the abundance of water $(\Sigma_\mathrm{H_2O}/\Sigma_\mathrm{g})(r,t)$ as shaped by pebble drift and radial advection and diffusion, water abundances in the disc surface layers may be considerably lower as strong UV irradiation by the host star (or the environment) and chemical reactions can process water molecules on timescales as low as $t_\mathrm{chem} = 10 - 100 \mathrm{~yr}$ \citep[][]{romero2024retrieval}. For the purposes of this work, we estimate the water abundance by comparing the local photo-dissociation timescale ($t_\mathrm{chem}$) to how quickly it is replenished from the midplane by vertical mixing, on timescales $t_\mathrm{z, mix} = 1/(\delta_\mathrm{vert}\Omega_\mathrm{K})$ \citep{semenov2011chemical}, which is $\lesssim 1000 \mathrm{~yr}$ inside the water snowline for our disc model.}

\AdrienRevision{We proceed as follows. For a region in the disc surface (for the purpose of this work, defined as above $\tau_\mathrm{20 \mu m}=1$), we approximate the water removal through local UV and chemical processing as}

\begin{equation}
\frac{d}{dt} \left(\frac{\rho_\mathrm{H_2O} }{\rho_g} \right)_\mathrm{UV} \approx - \left(\frac{\rho_\mathrm{H_2O} }{\rho_g}\right) \frac{1}{t_\mathrm{chem}}. 
\end{equation}

\AdrienRevision{Vertical mixing, however, will try to bring the local water abundance back to the (constant, on these timescales) column-integrated-one\footnote{\AdrienRevision{For columns outside the midplane snowline, a fraction of the water in the column is frozen out on grains. We compute the column-integrated water abundance $\Sigma_\mathrm{H_2O}/\Sigma_\mathrm{g}$ by integrating the gas and vapor from the snowline instead of the midplane.}}, on a vertical mixing timescale $t_\mathrm{z, mix}$, as}
\begin{equation}
\frac{d}{dt} \left(\frac{\rho_\mathrm{H_2O} }{\rho_g} \right)_\mathrm{mixing} \approx \left( \frac{\Sigma_\mathrm{H_2O}}{\Sigma_\mathrm{g}} \right)\frac{1}{t_\mathrm{z, mix}}.
\end{equation}
\AdrienRevision{Assuming these processes equilibrate on timescales shorter than radial transport ones and that $t_\mathrm{chem} < t_\mathrm{z, mix}$, we find}
\begin{equation}\label{eq:2D_water_new}
 \left(\frac{\rho_\mathrm{H_2O} }{\rho_g} \right)_\mathrm{surface} \approx \left( \frac{\Sigma_\mathrm{H_2O}}{\Sigma_\mathrm{g}} \right)\frac{t_\mathrm{chem}}{t_\mathrm{z, mix}},
\end{equation}
\AdrienRevision{such that the water vapor abundance in the disc surface layers is connected to the ratio between chemical processing and vertical mixing $t_\mathrm{chem}/t_\mathrm{z, mix} \leq 1$. Note that this approach does not include the impact of water losses occurring in the surface layers on the actual surface density evolution of vapor, as usually only a small fraction of the bulk vapor is present above the dust $\tau = 1$ surface (see Fig.~\ref{fig:f_obs}).}

\AdrienRevision{A detailed treatment of how $t_\mathrm{chem}$ varies with position, time and other factors is beyond the scope of this work. Instead, we consider two representative cases in the remainder of this work, one in which $t_\mathrm{chem} = 10 \mathrm{~yr}$ \citep[as typically $t_\mathrm{chem} = 10 - 100 \mathrm{~yr}$, see e.g.,][]{romero2024retrieval} and $t_\mathrm{z, mix} = 1/(\delta_\mathrm{vert}\Omega_\mathrm{K})$, and a second case where the ratio $t_\mathrm{chem}/t_\mathrm{z, mix} = 0.01$ is kept constant for all $r$.}

\section{Results}
\label{sec:results}

\subsection{Evolution of the bulk dust and water content}

\subsubsection{Dust coagulation and transport}
\label{sec:results_dust_coag}

\begin{figure*}
\begin{center}

\includegraphics[width=\textwidth]{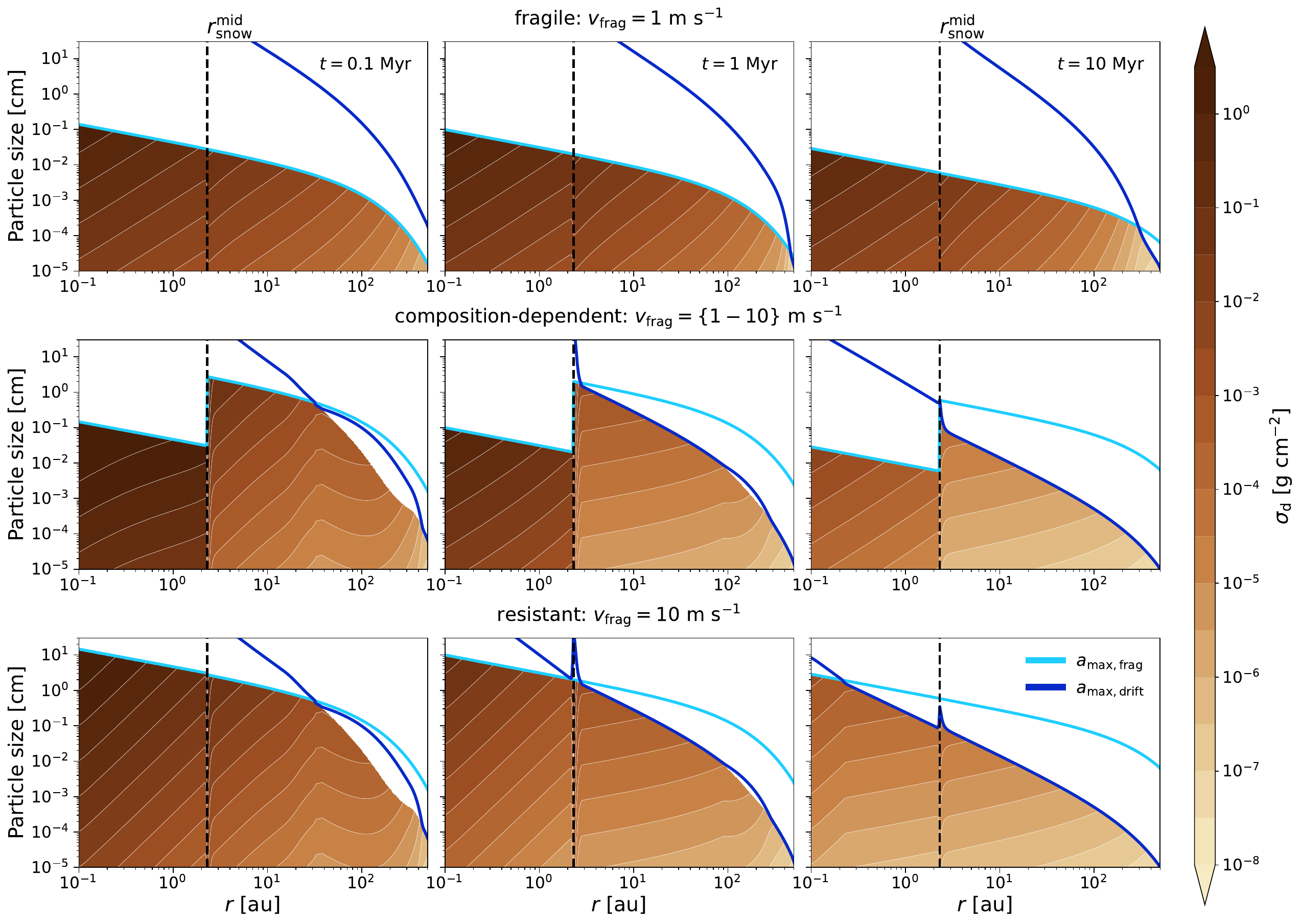}
    \caption{Vertically-integrated dust density distribution as a function of particle size for the fragile model (upper panels), composition-dependent model (middle panels), and resistant model (lower panels). The colour bar refers to the surface density of dust in a given size range at a given radius. The three panels represents different times of the simulation ($t = 0.1 \mathrm{~Myr}$, $t = 1 \mathrm{~Myr}$, and $t = 10\mathrm{~Myr}$). The vertical dotted black line indicates the midplane water snowline. The light and dark blue solid lines show estimates of the fragmentation and radial drift growth barrier, respectively.}
    \label{fig:1D_radial_dust_distribution}
\end{center}
\end{figure*}

\AdrienRevision{Focusing first on the evolution of the dust population,} we present in Fig.~\ref{fig:1D_radial_dust_distribution} the vertically-integrated dust density distribution as a function of particle size at different times ($t = 0.1$, $1$, and $10 \mathrm{~Myr}$) for the fragile (upper panels), composition-dependent (middle panels), and resistant dust model (lower panels). 

\AdrienRevision{The growth timescale is dependent on the dust densities, lower in the outer disc, such that the growth front propagates outwards (see e.g., middle and lower panels at $0.1 \mathrm{~Myr}$ outside $50 \mathrm{~au}$).} Particles coagulate to larger and larger aggregates, until their growth is halted either by the fragmentation barrier (light blue line) or the radial drift barrier (dark blue line). In the resistant and composition-dependent models, particles in the outer disc are more resistant to fragmentation, reaching particle sizes $a_\mathrm{max} \gtrsim 1 \mathrm{~cm}$, resulting in an efficient radial drift. After $t = 10 \mathrm{~Myr}$, there remains only a fraction of the initial solid reservoir. In the composition-dependent model, however, the lowered fragmentation velocity inside the snowline results in smaller particle sizes \citep[$a_\mathrm{max} \propto v_\mathrm{f}^2$, see][]{birnstiel2011dust} and weaker radial drift, accumulating solid material in the inner region\edited{, i.e., a traffic-jam \citep{banzatti2015direct, pinilla2016tunnel, pinilla2017dust}}. In the fragile dust model, particles remain below $1 \mathrm{~mm}$ and drift much more slowly. Even after $t = 10 \mathrm{~Myr}$, there is still a large fraction of the initial solid reservoir present in the disc.

\subsubsection{Pebble flux and bulk water vapor content}
\label{sec:results_flux}

\begin{figure}
\begin{center}

\includegraphics[width=\columnwidth]{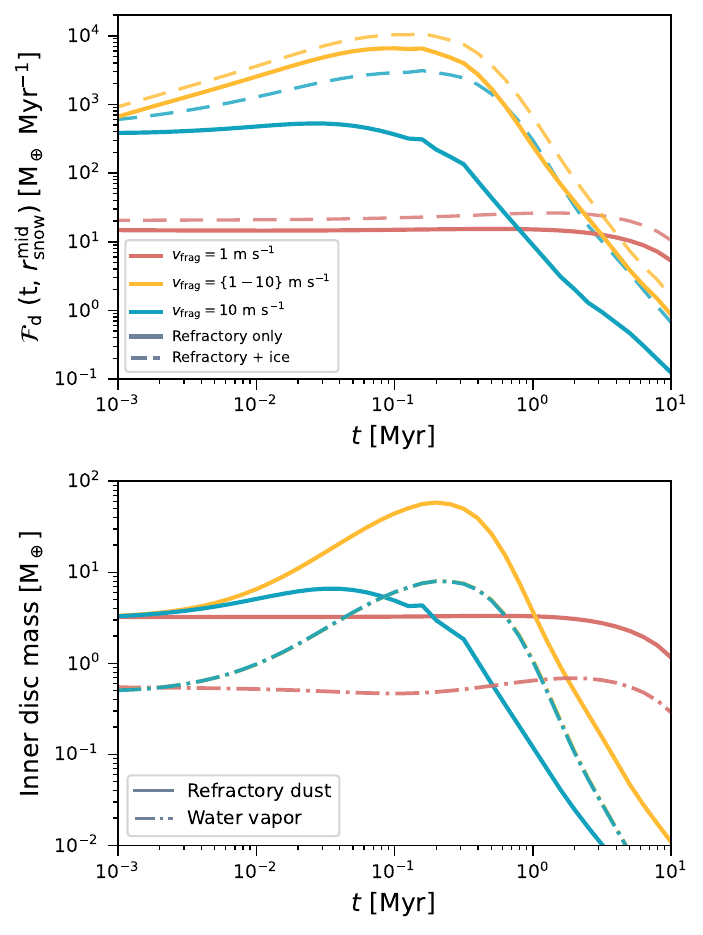}
    \caption{Mass flux of dust particles through the midplane water snowline $r_\mathrm{snow}^\mathrm{mid}$ \AdrienRevision{either including or excluding the water ice mass (upper panel) and inner disc dust and vapor mass (lower panel)} as a function of time for the three dust models.}
    \label{fig:pebble_flux}
\end{center}
\end{figure}

\AdrienRevision{As icy pebbles drift inward from the outer regions, they carry a mantle of water ice that sublimates once inside the snowline, leading in an enrichment of the bulk water vapor mass in the inner disc.} We present in Fig.~\ref{fig:pebble_flux} the pebble flux $\mathcal{F}_\mathrm{d}$ through the midplane water snowline $r_\mathrm{snow}^\mathrm{mid}$ as a function of time for the three dust models. Similarly to what we discussed in Section~\ref{sec:results_dust_coag}, particles in the fragile dust model are small, and the radial drift not sufficient to create a high pebble flux. It remains almost constant with time. In the two other models characterized by $v_\mathrm{frag} = 10 \mathrm{~m~s^{-1}}$ in the outer disc, the pebble flux is a few orders of magnitude larger. It follows a similar evolution to what is usually found in the literature \citep[e.g.,][]{drazkowska2021how}, with $\mathcal{F}_\mathrm{d}$ decreasing at later stage of evolution, once most of the disc solid material has drifted inward. \AdrienRevision{Differences between models can arise from differences between the interplay of dust and ice around the snowline. Most dramatically, the difference between the pebble flux including or excluding the water ice mass is due to the outward diffusion of vapor, that leads to water re-condensating and crossing multiple times the snowline, which increases the mass flux.}

\AdrienRevision{To illustrate the inner disc's enrichment in dust and water and compare to previous studies, we show} in the lower panel in Fig.~\ref{fig:pebble_flux} the total dust \AdrienRevision{and water vapor mass} inside the midplane snowline. In our disc model, we initially find a few ${\sim}\mathrm{~M_\oplus}$ of dust particles \AdrienRevision{and about $20\%$ of this mass in water vapor}. The two models with constant $v_\mathrm{frag}$ (fragile and resistant models) display an almost constant dust mass that decays at larger time as the outer disc is emptying out of dust particles. It decays faster for the resistant dust model, because larger particles drift through the disc more rapidly. In the composition-dependent model, the \AdrienRevision{lowered fragmentation velocity inside the snowline leads in an accumulation of dust (i.e., traffic jam effect) peaking above ${\sim}50 \mathrm{~M_\oplus}$ around $t \sim 0.3 \mathrm{~Myr}$. The total water vapor mass follows a similar evolution to previous works \citep[][]{kalyaan2021linking, kalyaan2023effect}, initially increasing due to ongoing delivery by drifting pebbles, before decreasing at later stages as the pebble flux drops and water vapor is advected onto the central star. The total mass of water vapor can become as high as $10 \mathrm{~M_\oplus}$ in the composition-dependent and resistant models, where the radial pebble flux is higher.}

\subsection{Vertical dust and vapor densities and the evolution of the $\tau=1$ surface}

\begin{figure*}
\begin{center}

\includegraphics[width=\textwidth]{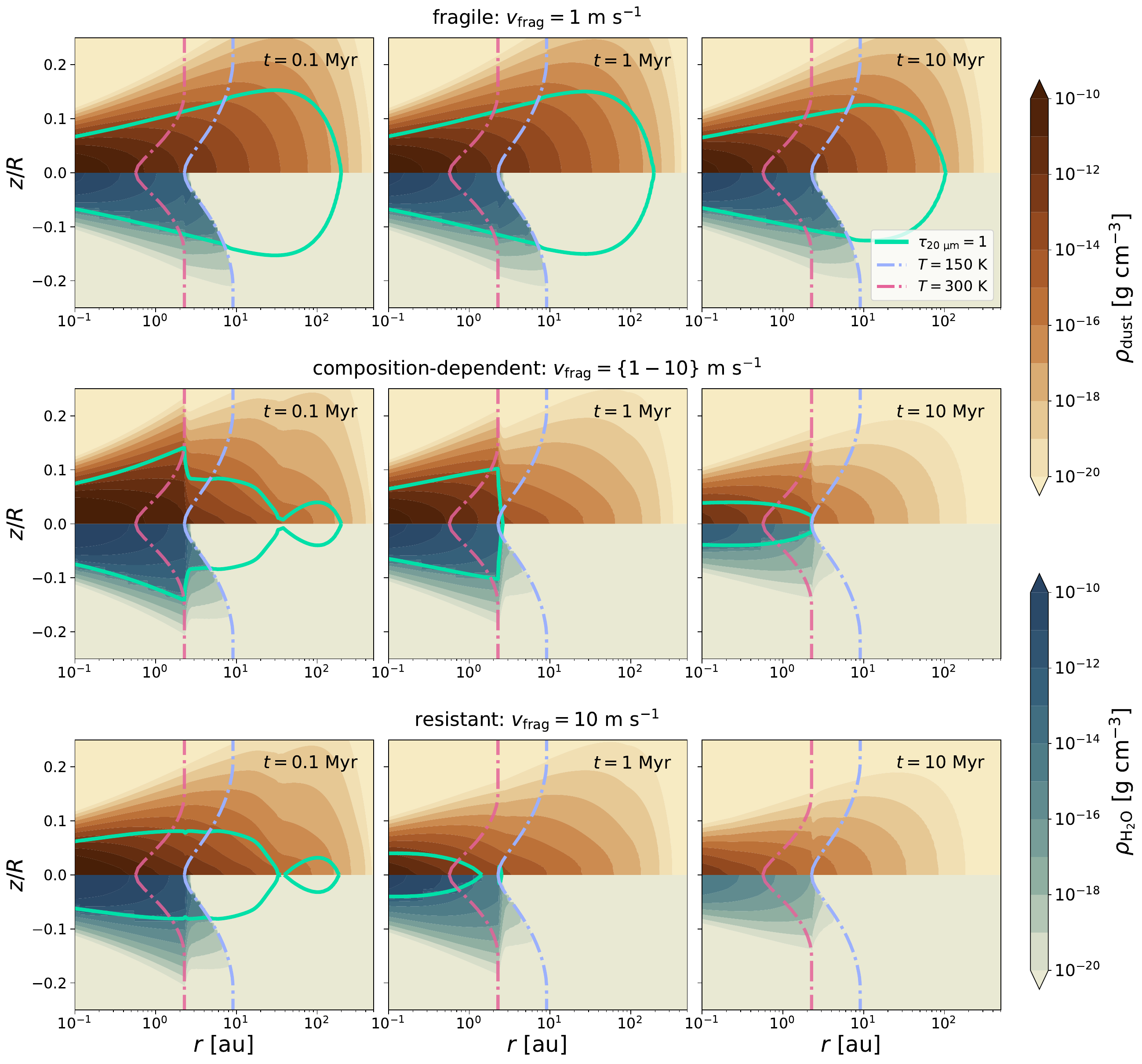}
    \caption{Vertical dust \AdrienRevision{(in the $z > 0$ parts)} and water vapor \AdrienRevision{(in the $z < 0$ parts, assuming $t_\mathrm{chem}/t_\mathrm{z, mix} = 0.01$)} density distribution for the fragile model (upper panels), composition-dependent model (middle panels), and resistant model (lower panels) at different times ($t = 0.1 \mathrm{~Myr}$, $t = 1 \mathrm{~Myr}$, and $t = 10 \mathrm{~Myr}$). We obtained the vertical structure from the outputs of \texttt{chemcomp}, assuming vertical hydrostatic equilibrium (Equation~\ref{eq:dust_density}). The \AdrienRevision{upper} colour bar indicates the dust density $\rho_\mathrm{d}$ for all particle sizes. The green solid line indicates the height of the $\tau_\nu = 1$ surface at $20 \mathrm{~\mu m}$, representing a typical wavelengths observed by JWST. \AdrienRevision{The dotted lines specify the temperature contour associated with $T=150\mathrm{~K}$ (i.e., the water snowline) and $T=300\mathrm{~K}$ (used in Fig.~\ref{fig:FigY_obs_water_T_componant} to define the cold and warm temperature components), obtained from Equation~\ref{eq:2D_gas_temperature}.}}
    \label{fig:2D_dust_distribution}
\end{center}
\end{figure*}

\AdrienRevision{Moving towards a 2D picture,} we now present in Fig.~\ref{fig:2D_dust_distribution} \AdrienRevision{(in the $z > 0$ parts)} the vertical dust density distribution, assuming a (grain-size dependent) equilibrium between settling and turbulent mixing (see Sect.~\ref{sec:vertical_distrib_methods}). The colormap indicates the dust density for all particle sizes. We also add the 2D water snowline profile as computed with Equation~\ref{eq:2D_gas_temperature} along with the $\tau = 1$ surface of the dust continuum at $20 \mathrm{~\mu m}$, representing a \edited{typical wavelengths observed by} JWST\footnote{\edited{JWST-MIRI ranges from $5$ to $28 \mathrm{~\mu m}$. We choose $20 \mathrm{~\mu m}$, towards the upper end of the instrument range, because that is where cold water lines have been predominantly observed \citep{banzatti2024atlas}. \AdrienRevision{The dust opacity does not vary significantly across these wavelengths, so our results should remain similar.}}}. \edited{Note that in this paper, we do not model specific spectral lines. The choice of a wavelength is necessary only due to the wavelength-dependence of the dust optical depth and dust opacity.} \edited{The optical depth is computed with Equation~\ref{eq:optical_depth}, taking into account the evolving size distribution at every $(r,z)$.}

\AdrienRevision{We see that except for the fragile dust model, for which the inward transport is low, the height of the $\tau = 1$ surface varies significantly with time. At earlier stages, the disc is optically thick up to hundreds of au, decreasing with time. After $t = 10 \mathrm{~Myr}$, an important fraction of the total dust mass has been accreted onto the central star for the composition-dependent and resistant models (for which icy pebbles drift efficiently). The disc becomes optically thin at $20 \mathrm{~\mu m}$ for the resistant model, while due to the traffic-jam effect, the composition-dependent model maintains an optically thick region at $20 \mathrm{~\mu m}$ up to $2 \mathrm{~au}$. The entire midplane reservoir of vapor becomes observable only once the disc becomes optically thin. At this stage, the only way to hide water in the column from view is for it to be frozen on dust grains.}

\begin{figure}
\begin{center}
\includegraphics[width=\columnwidth]{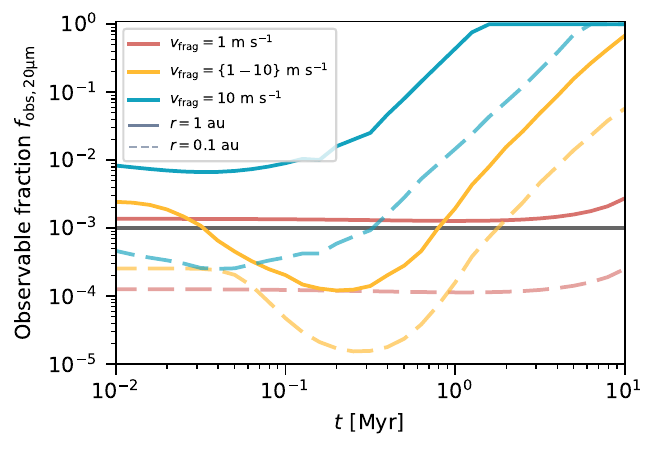}
    \caption{\AdrienRevision{Observable fraction of gaseous material $f_\mathrm{obs} = \Sigma_\mathrm{g, obs}/(\Sigma_\mathrm{g, obs} + \Sigma_\mathrm{g, hidden})$ relative to the total gas content in the vertical column at $20 \mathrm{~\mu m}$. The hidden reservoir $\Sigma_\mathrm{g, hidden}$ and observable reservoir $\Sigma_\mathrm{g, obs}$ are obtained by vertically integrating the gas content from the midplane to the $\tau_\mathrm{~20 \mu m} = 1$, and from this surface to the top of the disc atmosphere, respectively. In the absence of UV and chemical processing in the disc atmosphere, water vapor is observable in the same fraction. The horizontal line indicates for comparison the value used by \citet{romero2024retrieval}.}}
    \label{fig:f_obs}
\end{center}
\end{figure}

\AdrienRevision{To highlight the time evolution, we show in Fig.~\ref{fig:f_obs} the observable fraction of gaseous material above the $\tau_\mathrm{~20\mu m} = 1$ surface compared to the total mass in the column, at $0.1$ and $1 \mathrm{~au}$ for the different dust models. It reaches as low as $10^{-5}$ in the composition-dependent case (a result of the high dust densities following the traffic-jam inside the snowline), and tends to increase with time after $t = 0.3 \mathrm{~Myr}$ until it reaches $f_\mathrm{obs, 20\mu m} = 1$, where the column is optically thin and the whole column is observable. The observable fraction decreases closer to the star, by about one order of magnitude between $1 \mathrm{~au}$  and $0.1 \mathrm{~au}$.}

Overall, we see that \AdrienRevision{(1) the coagulation, fragmentation, and transport of dust particles cause the $\tau_\nu=1$ layer to shift up/down with time, resulting in a smaller/larger fraction of the disc material to be visible, and (2) the exact behavior depends sensitively on the assumed dust model.}

\subsection{2D maps of the water vapor abundance}

\AdrienRevision{To visualize and summarize how the various processes mentioned so far interact, we also plot in Fig.~\ref{fig:2D_dust_distribution} (in the $z < 0$ parts) the 2D water vapor density $\rho_\mathrm{H_2O}$ as obtained using the methodology outlined in Sect.~\ref{sec:methods_2D}, assuming $t_\mathrm{chem}/t_\mathrm{z, mix} = 0.01$. The observed behavior can be understood as follows:
(1) Near the midplane and inside the midplane snowline ($r < r_\mathrm{snow}^\mathrm{mid}$), we find high densities of water vapor. Though water is released at the snowline by inward-drifting icy pebbles, the plume of vapor rapidly evolves from a local overdensity to distribute throughout the inner disc \citep[see also][]{cuzzi2004material}. The abundance of vapor relative to the bulk gas is typically constant\footnote{These results are different than what can be seen at snowlines in the outer regions, e.g., at the CO snowline \citep[e.g.,][]{stammler2017redistribution}, where a local pile-up of gaseous material may appear due to much longer diffusion timescales.} (as a function $r$), as in the inner disc the radial and vertical mixing timescales are much shorter than the timescale at which the pebble flux evolves (see Fig.~\ref{fig:pebble_flux}). (2) In the surface layers overhead ($z > z(\tau_\mathrm{20 \mu m} = 1)$), the abundance drops significantly due to the assumed rapid local processing ($t_\mathrm{chem}/t_\mathrm{z, mix} = 0.01$, see Sect.~\ref{sec:UV_loss_methods}). (3) Just outside the midplane snowline, the midplane contains no vapor (water is frozen out on grains) but water vapor still exists in the warmer ($T > 150 \mathrm{~K}$) surface layers (see Sect.~\ref{sec:2D_temperature}). The density of vapor in the surface layers is decreased compared to the inner disc due to a large fraction of the column's water being frozen out on large dust grains. (4) Outside $r\sim10\mathrm{~au}$, the water snowline reaches the disc surface and water in the entire column is present predominantly as ice.}

\subsection{Evolution of the observable water vapor}

\subsubsection{Temporal evolution of the observable water vapor}
\label{sec:results_N_H2O_vs_time}

Finally, we can investigate how the column density of observable water vapor changes with time. The observable column density is calculated from Equation~\ref{eq:observable_column_density}. Given the focus of this work on JWST, we focus again on the impact of dust extinction at $20 \mathrm{~\mu m}$. 

\begin{figure}
\begin{center}

\includegraphics[width=\columnwidth]{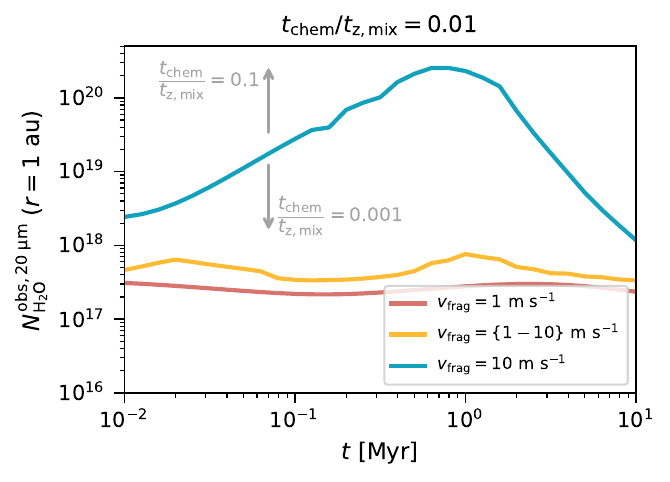}
    \caption{\AdrienRevision{Observable column density of water vapor above the $\tau_\mathrm{20 \mu m} = 1$ surface of the dust continuum at $1 \mathrm{~au}$, as a function of time for three dust models, assuming $t_\mathrm{chem}/t_\mathrm{z, mix} = 0.01$ (efficient UV and chemical processing of vapor)}.}
    \label{fig:Fig4_column_density_vs_time}
\end{center}
\end{figure}

We present in Fig.~\ref{fig:Fig4_column_density_vs_time} the temporal evolution of the column density of observable water vapor at $1 \mathrm{~au}$ at $20 \mathrm{~\mu m}$ for the three dust models\AdrienRevision{, assuming $t_\mathrm{chem}/t_\mathrm{z, mix} = 0.01$ (efficient UV and chemical processing)}. This location is inside the midplane water snowline, and is characterized by $T = 225 \mathrm{~K}$ in the midplane, and $T= 450 \mathrm{~K}$ in the disc atmosphere. 

\AdrienRevision{For the $v_\mathrm{frag} = 1 \mathrm{~m~s^{-1}}$ case, where inward drift is inefficient, the observable vapor reservoir is almost constant with values $N_\mathrm{H_2O}^\mathrm{obs} \sim 10^{17} \mathrm{~cm^{-2}}$. The column density depends proportionally on the ratio between UV/chemical processing and replenishment by vertical mixing (see Eq.~\ref{eq:2D_water_new}), such that in the absence of chemical processing, we would find $N_\mathrm{H_2O}^\mathrm{obs} \sim 10^{19} \mathrm{~cm^{-2}}$. For the $v_\mathrm{frag} = 10 \mathrm{~m~s^{-1}}$ case, the evolution of the observable reservoir follows a bell-shaped curve, varying by several orders of magnitude in time between $N_\mathrm{H_2O}^\mathrm{obs} \sim 10^{18} \mathrm{~cm^{-2}}$ and $N_\mathrm{H_2O}^\mathrm{obs} \sim 10^{20} \mathrm{~cm^{-2}}$, when using $t_\mathrm{chem}/t_\mathrm{z, mix} = 0.01$. Interestingly, the peak of the observable vapor reservoir at $1 \mathrm{~au}$ peaks at $1 \mathrm{~Myr}$, long after the actual peak of the bulk vapor mass in the inner disc ($0.3 \mathrm{~Myr}$, see Fig.~\ref{fig:pebble_flux}).} \AdrienRevision{Last, in the composition-dependent model, the accumulation of dust compensates for the active delivery of water vapor by icy pebbles. It results in an almost constant observable reservoir, just a few factors higher than for the fragile model in which the inward transport of pebbles is inefficient. It means that despite of the efficient delivery of water vapor to the inner regions ($\mathcal{F}_\mathrm{d} > 100 \mathrm{~M_\oplus ~Myr^{-1}}$, see Fig.~\ref{fig:pebble_flux}), the observable reservoir remains constant, to a value set by the interplay of UV/chemical processing and vertical mixing. Water vapor is effectively \AdrienFinalRev{smuggled} unnoticed.}

Overall, the column density of observable water in the inner disc is sensitive both to the actual flux of ice into the inner disc \emph{and} to the behaviour of dust inside the snowline. In the traffic jam scenario, the pile-up of (small) dust in the inner disc regions is so dramatic that it overshadows the delivery of water, and the column density of (observable) \AdrienRevision{water remains constant, even in the period in which the majority of volatiles is delivered (see Fig.~\ref{fig:pebble_flux}).}

\subsubsection{Radial dependence of $N_\mathrm{H_2O}$}
\label{sec:results_N_H2O_vs_rad}

\edited{As mentioned in Section~\ref{sec:intro}, \citet{banzatti2023jwst} recently found that a single temperature slab model cannot fully reproduce water vapor emission observed with JWST. Therefore, there is an ongoing effort to fit JWST infrared spectra rather with multiple temperature components \citep[e.g.,][]{pontoppidan2024high, temmink2024minds}} or even power-laws \AdrienRevision{\citep[][]{romero2024retrieval, kaeufer2024bayesian}}, in which the presence of pebble drift is hypothesised to manifest itself in particular as a "cold" water reservoir, with $T < 400\mathrm{~K}$, and emitting \edited{radii} typically $<10 \mathrm{~au}$. It thus becomes interesting to plot not only the temporal evolution of the observable column density at a fixed radius (see Fig.~\ref{fig:Fig4_column_density_vs_time}), but also its radial profile.

\begin{figure*}
\begin{center}
\includegraphics[width=\textwidth]{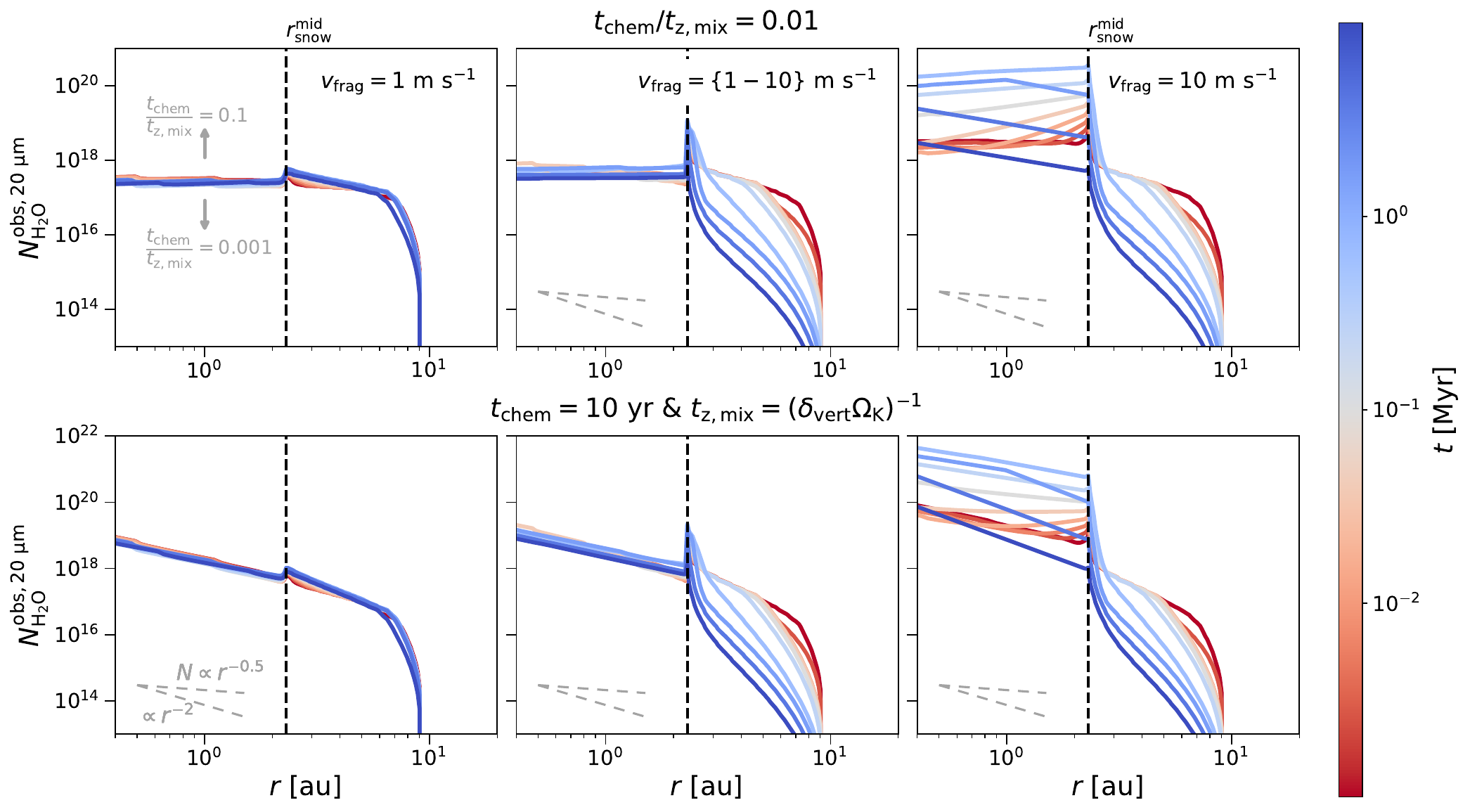}
    \caption{\AdrienRevision{Column density of observable water vapor above the $\tau_\mathrm{20 \mu m} = 1$ surface of the dust continuum as a function of radius for different times. We include the three dust models: fragile dust (left panels), composition-dependent (middle panels), and resistant dust (right panels), assuming $t_\mathrm{chem}/t_\mathrm{z, mix} = 0.01$ (efficient UV and chemical processing of vapor). Similarly to Fig.~\ref{fig:Fig4_column_density_vs_time}, the column density is proportional to $t_\mathrm{chem}/t_\mathrm{z, mix}$, and may be increased by two orders of magnitude in the absence of UV and chemical processing. The dashed vertical line indicates the position of the midplane water snowline, while the grey lines in the lower left inset represent typical slopes used to fit water vapor radial profile measured with JWST \citep{romero2024retrieval}.}}
    \label{fig:Fig5_column_density_vs_radius}
\end{center}
\end{figure*}

We present in Fig.~\ref{fig:Fig5_column_density_vs_radius} the radial profile of the observable column density of water vapor at $20 \mathrm{~\mu m}$ for the three dust models, \AdrienRevision{including the impact of UV and chemical processing on the vapor abundances in the upper disc layers. As mentioned in Sect.~\ref{sec:UV_loss_methods}, we consider a fixed ratio $t_\mathrm{chem}/t_\mathrm{z, mix} = 0.01$ (upper panels) and a fixed $t_\mathrm{chem} = 10 \mathrm{~yr}$ with a mixing timescale varying radially following its definition (lower panels).} 

In the fragile dust model (left panels in Fig.~\ref{fig:Fig5_column_density_vs_radius}), the radial profile inside $\sim 6 \mathrm{~au}$ is governed by the $\tau = 1$ surface, resulting in an almost flat profile. We explain that flatness due to the compensation between the steadily increasing height of the $\tau = 1$ with distance to the star (see upper panels in Fig.~\ref{fig:2D_dust_distribution}) and the increasing gas scale-height (as $h_\mathrm{g} \propto r^{5/4}$, see Section~\ref{sec:methods}). Between $6$ and $9 \mathrm{~au}$, a transition occurs as the $\tau = 1$ surface passes below the water snowline. \AdrienRevision{In consequence, the radial profile of observable vapor decreases more steeply, following} the geometry of the snowline. Outside ${\sim}9 \mathrm{~au}$, the column density sharply drops as the temperature is below $150 \mathrm{~K}$ \AdrienRevision{even in the disc atmosphere. Similarly to what we found earlier (at $1 \mathrm{~au}$ in Fig.~\ref{fig:Fig5_column_density_vs_radius}), the radial profile of observable vapor is almost constant with time.}

\AdrienRevision{In the resistant dust model (right panel in Fig.~\ref{fig:Fig5_column_density_vs_radius}), the inner disc dust content is cleared more rapidly, such that the $\tau = 1$ surface is lower and the inner disc becomes optically thin more rapidly. Where the observable water column is limited by the $\tau = 1$ surface (typically for $r < r_\mathrm{snow}^\mathrm{mid}$), it results in column densities several orders of magnitude higher than the other models. At later stages ($t > 1 \mathrm{~Myr}$, see Fig.~\ref{fig:f_obs}), the inner disc becomes optically thin and the whole reservoir is observable. The radial profile then follows the gas density profile. As for the fragile dust model, the geometry of the snowline dictates the radial profile of observable vapor in the colder outer parts. The observable column densities are lower due to the large fraction of water in the column being stored as ice.}

In the composition-dependent model (middle panels in Fig.~\ref{fig:Fig5_column_density_vs_radius}), \AdrienRevision{the radial profile is a combination of the two other models. Outside the midplane water snowline, it is similar to the resistant dust model, where the snowline geometry governs the radial profile of observable vapor. Inside the midplane water snowline, the accumulation of dust hides a large fraction of vapor, leading to an almost flat - temporally constant - profile, as in the model where inward transport is inefficient ( $v_\mathrm{frag} = 1 \mathrm{~m~s^{-1}}$, left panels). Comparing to the resistant model, the observable column density for $r < r_\mathrm{snow}^\mathrm{mid}$ decreases by up to two orders of magnitude in the presence of a dust accumulation, even though the total water vapor mass is identical in both models (see Fig.~\ref{fig:pebble_flux}).}

\AdrienRevision{Concerning UV and chemical processing, we find similar results to Fig.~\ref{fig:Fig4_column_density_vs_time} when using a fixed $t_\mathrm{chem}/t_\mathrm{z, mix}$ ratio, with the observable column density being directly proportional. When $t_\mathrm{chem}$ is fixed but $t_\mathrm{z, mix} = (\delta_\mathrm{vert}\Omega_\mathrm{K})^{-1}$, the radial profile displays a slightly steeper slope as the impact of UV/chemical processing on water abundances depends on the radial distance. In the very inner regions, $t_\mathrm{z, mix}$ is small and the $t_\mathrm{chem}/t_\mathrm{z, mix}$ ratio closer to unity, where UV/chemical processing has little impact on the water abundances.}

In the end, we find that $N_\mathrm{H_2O}^\mathrm{obs}$ is highly variable, depending on the intensity of pebble drift, the presence or absence of a traffic-jam, and \AdrienRevision{the efficiency of UV and chemical processing}. The radial profile $N_\mathrm{H_2O}^\mathrm{obs}(r)$ over time \AdrienRevision{, as shaped by the evolution of both the water and dust abundance,} displays different shapes (flat, power-law like) and spans across multiple orders of magnitude, from $10^{14}$ to $10^{20} \mathrm{~cm^{-2}}$.

\section{Discussion}
\label{sec:discussion_conclusion}

\subsection{Comparisons with observational trends}

Overall, the observable column density of water vapor spans many orders of magnitude through the disc lifetime, and the radial profile may have different shapes. Low fragmentation velocities (weak drift) tend to give fairly flat radial profiles for $N_\mathrm{H_2O}^\mathrm{obs}$, where the actual value depends on the drift rate and \AdrienRevision{$t_\mathrm{chem}$}. High fragmentation velocities (fast drift) result in more centrally peaked radial profiles for $N_\mathrm{H_2O}^\mathrm{obs}$, due to the greater influence of the snowline geometry. The traffic-jam effect (uniquely) results in a low value inside the snowline combined with a \AdrienRevision{steep radially decreasing trend} outside it. It is difficult to draw conclusions about the fragmentation velocity of dust particles by comparing our results obtained with a single disc model with population-level trends observed with Spitzer \citep[e.g.,][]{carr2008organics} or JWST \citep[e.g.,][]{banzatti2023jwst}. However, it is interesting to note that the flat column density profile obtained with the low fragmentation velocity case could explain why one-component (i.e., single temperature and column density) slab models can accurately reproduce many features in several cases \citep[e.g.,][]{carr2008organics, grant2023minds, banzatti2023jwst}. Nevertheless, recent studies that have gone beyond a single 0D slab model, e.g., fitting (tapered) power-laws \AdrienRevision{\citep[][]{romero2024retrieval, kaeufer2024bayesian}}, seem to favour decreasing column density with radius, which matches the high fragmentation velocity cases (see \ref{fig:Fig5_column_density_vs_radius}). 

\subsection{Cold water mass as a pebble drift tracer}

\begin{figure*}
\begin{center}
\includegraphics[width=\textwidth]{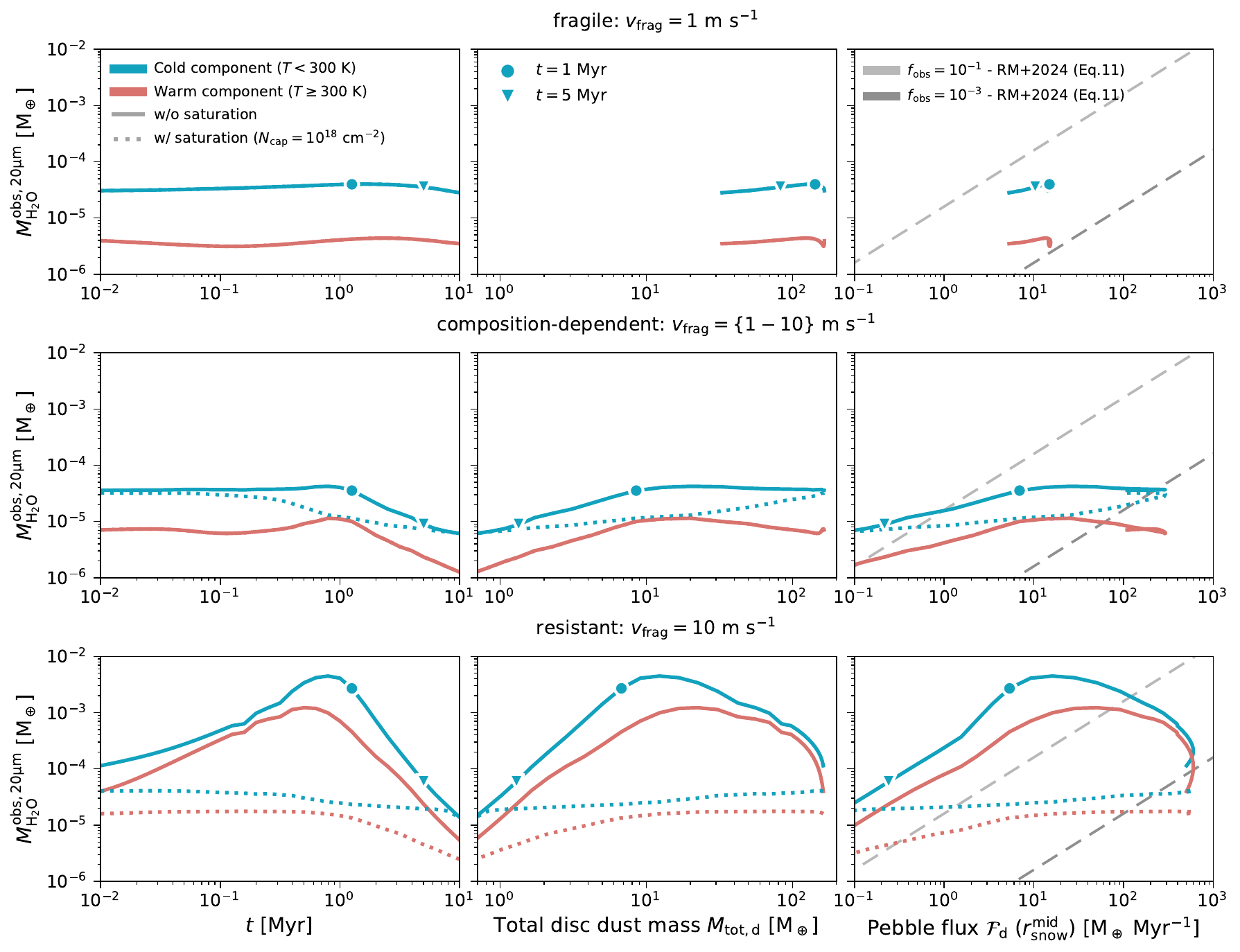}
    \caption{\AdrienRevision{Total observable water vapor mass $M_\mathrm{H_2O}$ at $20\mathrm{~\mu m}$, as a function of time (left panels), total remaining disc dust mass (middle panels) and refractory pebble flux through $r_\mathrm
    {snow}^\mathrm{mid}$ (right panels) for the three dust models assuming $t_\mathrm{chem}/t_\mathrm{z, mix} = 0.01$ (efficient UV and chemical processing of vapor). The observable vapor content is divided in two temperature components: a cold reservoir ($T < 300 \mathrm{~K}$) and a warm reservoir ($T \geq 300 \mathrm{~K}$), similarly to what is found from JWST observations \citep[e.g.,][]{banzatti2023jwst, banzatti2024atlas}. The dashed lines account for the fact that if the column density at a given radius is too high (typically $N_\mathrm{cap} > 10^{18}$), water emission may become optically thick and a higher fraction of vapor may be hidden, additionally hiding temporal evolution. Blue markers indicate $t = 1 \mathrm{~Myr}$ and $t = 5 \mathrm{~Myr}$ to visualize in which directions the curves evolve with time. We include in the right panels the relation between pebble flux and water vapor mass as used in \citet{romero2024retrieval} (Eq.~11, where they assumed $\eta = (f_\mathrm{obs})^{-1} = 10^3$).}}
    \label{fig:FigY_obs_water_T_componant}
\end{center}
\end{figure*}

\AdrienRevision{As detailed in Sect.~\ref{sec:intro}, the mass of cold water vapour, inferred from JWST/MIRI spectra, has emerged as a potential direct tracer of ongoing pebble flux \citep[][]{banzatti2023jwst, banzatti2024atlas}. Indeed, in \citet{romero2024retrieval} a linear relation between $M_\mathrm{H_2O, cold}$ and $F_\mathrm{peb}(r_\mathrm{snow}^\mathrm{mid})$ is assumed (their Eq.~11). The models developed here, in which the amount of cold water depends on variety of processes all operating on different timescales, allow us to look at this link in a new way. Fig.~\ref{fig:FigY_obs_water_T_componant} presents the total integrated water mass $M_\mathrm{H_2O}$ for the cold ($T < 300 \mathrm{~K}$) and hot ($T \geq 300 \mathrm{~K}$) water components (for $t_\mathrm{chem}/t_\mathrm{z, mix} = 0.01$) as a function of time (left column), remaining dust mass (middle), and current refractory pebble flux through the snowline (right column). For comparison, cold water masses inferred from JWST/MIRI spectra are typically around $\lesssim 0.1 - 10 \mathrm{~\mu M_\oplus}$ \citep[][]{romero2024retrieval, banzatti2024atlas}.}

\AdrienRevision{We focus first on the solid lines in Fig.~\ref{fig:FigY_obs_water_T_componant}, which depict the total warm and cold mass reservoirs above the layer where the dust becomes optically thick, i.e., essentially a radial integral over $N_\mathrm{H_2O}^\mathrm{obs}(r)$. The left column highlights that the observable water reservoirs increase primarily in the resistant scenario (bottom row), where observable warm and cold component masses peak between 0.1-1 Myr, comparable to when the entire inner disc becomes enriched in water vapour (see Fig.~\ref{fig:pebble_flux}). In the fragile model little change is observed, and in the composition-dependent case (middle row) only a (very) small rise is seen around 1 Myr followed by a slow decrease. We attribute the lack of a clear peak to the increase in dust in the inner disc as a result of the traffic jam.}

\AdrienRevision{A more useful comparison from an observational perspective may be the middle column, where we show the same water masses but now against the total dust and pebble mass remaining in the system. Dust masses in our model are monotonically decreasing, such that discs now evolve from right to left, but the speed at which dust mass is lost depends on the particle size and hence fragmentation velocity \citep[e.g.,][]{birnstiel2009dust, birnstiel2012simple}, leading to different behaviour between the three dust models. Only in the resistant dust model do we find an increase in the observable water mass with decreasing disc dust mass \citep[][]{najita2013hcn}, at least for $t < 1 \mathrm{~Myr}$.}

\AdrienRevision{Finally, the right column in Fig.~\ref{fig:FigY_obs_water_T_componant} shows the cold (and warm) water masses against the (instantaneous) refractory pebble mass flux crossing the water snowline. It is generally assumed that there exists a linear relationship between the observable water vapor mass and pebble flux \citep[e.g.,][]{romero2024retrieval}. However, except for the resistant model for $t > 1 \mathrm{~Myr}$ that displays a linear trend, we can see that it is generally a more complex relation. For the composition-dependent model, where a traffic-jam is present, a broad range of pebble flux could be associated to $M_\mathrm{H_2O, cold} \sim 10^{-5} \mathrm{~M_\oplus}$. One reason for the deviation from a simple $\mathcal{F}_\mathrm{d} \propto M_\mathrm{H_2O}$ relation is the fact that $f_\mathrm{obs}$ has become a complex function of time (see Fig.~\ref{fig:f_obs}), but another one is that in our model the amount of (observable) water in the inner disc depends not only on the current pebble flux but also its past evolution and on how water vapor distributed radially and vertically (Sect.~\ref{sec:methods}).}

\AdrienRevision{So far, we have assumed that all water vapour located above $z(\tau_\mathrm{20 \mu m} = 1)$ contributes to $M_\mathrm{H_2O, cold}$ and $M_\mathrm{H_2O, warm}$. In practice, individual emission lines themselves will become optically thick if the water abundance is high. To illustrate the potential impact, the dashed lines in Fig.~\ref{fig:FigY_obs_water_T_componant} show what the observable water reservoirs would be if we cap, at every radius, the water column at $10^{18} \mathrm{~cm^{-2}}$ when integrating over $N_\mathrm{H_2O}^\mathrm{obs}(r)$\footnote{This value was chosen because for typical slab models presented in \citet{banzatti2024atlas}, the majority of lines that would be detected in the MIRI spectrum become significantly optically thick between $10^{17}-10^{18}\mathrm{~cm^{-2}}$, however, the exact behaviour will vary from line to line.}. The cap has a significant impact for the resistant case, for which removing the highest column densities removes most of the variation seen with time. For the cold water, the reason for this is that the majority of this reservoir is found in a narrow, high column density region just outside the snowline (see Fig.~\ref{fig:Fig5_column_density_vs_radius}). A full 2D advection/diffusion calculation (see next Section) is needed to investigate whether this narrow region can persist, despite the large concentration gradient in the disc surface, or whether the cold water will spread further out and thus become more optically thin. Either way, retrievals frequently return column densities up to ${\sim}10^{19}\mathrm{~cm^{-2}}$ \citep[e.g.,][]{banzatti2024atlas}, suggesting radial variations as those presented in Fig.~\ref{fig:Fig5_column_density_vs_radius} can be detected. Looking further ahead, future far-IR missions, like PRIMA \citep[covering wavelengths in the $35-200 \mathrm{~\mu m}$ range][]{pontoppidan2023prima}, could investigate many water lines with lower upper energy levels that are sensitive to cold water at larger - more optically thin - radii, which would help in constraining the radial profile and temporal variation of the observable vapor reservoir.}

\subsection{Limitations and future work}
\label{sec:lim_futur_work}

\AdrienRevision{Our approach constitutes a next step in considering the impact of high pebble flux on the delivery of both dust and water vapor to the inner regions, and there are a few aspects that could be improved for future works.}

\AdrienRevision{First, we use a simplified approach for the treatment of UV and chemical processing, either with a constant $t_\mathrm{chem}$, or a constant ratio $t_\mathrm{chem}/t_\mathrm{z, mix}$. Moreover, we do not include the feedback of the losses of vapor mass on the surface density evolution. While that is a satisfying assumption for most of the disc lifetime, where only a small fraction of the vapor reservoir gets photo-dissociated ($f_\mathrm{obs} << 1$, see Fig.~\ref{fig:f_obs}), it could have an impact at later stages at the disc becomes more optically thin.} Thermo-chemical models like \texttt{ProDiMo} \citep{woitke2009radiation} or \texttt{DALI} \citep{bruderer2012warm} are suited \AdrienRevision{to better estimate UV and chemical processing as a complex function of time, height and radial distance}, but they do not include water or dust transport, hence miss the processes responsible for the orders of magnitude variations seen in for example Fig.~\ref{fig:Fig5_column_density_vs_radius}. A combined 2D study where the results of dynamics and pebble drift are included in thermo-chemical models would offer a solution \citep[see][for recent attempts in combining chemistry with dust evolution]{facchini2017different, woitke2022mixing, clepper2022chemical}. 

\AdrienRevision{Second, our 1+1D approach treats advection and diffusion in a vertically-integrated way (through the \texttt{chemcomp} computation) rather than treating these processes in full 2D \citep[e.g.,][]{ciesla2009two, krijt2020co}. This may lead to us not capturing diffusion of water vapour or small dust grains in the surface layers accurately when the radial abundance gradients in the disc surface differ from the ones closer to the midplane. When treated fully, these effects may for example smooth out the transition at $r_\mathrm{snow}^\mathrm{mid}$ seen in Fig.~\ref{fig:Fig5_column_density_vs_radius} as vapor delivered at the midplane snowline could diffuse outward above the snowline. Note, however, that \citet{ciesla2009two} finds that in their 2D transport models, most outward transport should occur close to the midplane. We also do not include the 2D shape of the snowline on the variable fragmentation velocity in the composition-dependent model. Nonetheless, this would likely have a small impact as the largest pebbles - those of which are close to the fragmentation barrier - are typically found in the midplane where water is in ice form (at the snowline, $h_\mathrm{peb}/r < 0.02$).}

\edited{Finally, we employ static, non-evolving temperature profile (see Equation~\ref{eq:2D_gas_temperature}), and hence do not consider the feedback of the evolving dust opacities (e.g., due to drift and dust growth) on temperature. Such feedback may result in radial and vertical shifts of the water snowline with time \citep[e.g.,][]{kondo2023roles, robinson2024cudisc}.}

We have focused on modeling a single set of initial disc conditions (e.g., $M_\mathrm{disc} = 0.05 \mathrm{~M_\odot}$, $\alpha=10^{-3}$), exploring the effects of the assumed dust model (high, low, or mixed fragmentation velocity). \AdrienFinalRev{The next step will be to perform a greater exploration of the disc parameter space, e.g., including different turbulence levels or initial disc mass and radius, to study how that affects the evolution of $N_\mathrm{H_2O}^\mathrm{obs}(r)$ over time, and the possibility of smuggling water unnoticed.} With a larger parameter space covered, it will be possible to investigate how $N_\mathrm{H_2O}^\mathrm{obs}$ correlate with \edited{stellar mass} \AdrienRevision{\citep[][]{pascucci2009different, pascucci2013atomic}}, or the underlying dust disc \AdrienRevision{mass/radius \citep{najita2013hcn, banzatti2020hints, banzatti2023jwst}}. \AdrienRevision{Two key ingredients that will also have to be included in future works are: (1) the presence of outer substructures, and their role in slowing down drift and allowing discs to maintain a large size \citep{kalyaan2021linking, banzatti2023jwst, xie2023water}, and (2) planetesimal formation, that may capture a fraction of the drifting ice and dust \citep[e.g.,][]{cuzzi2004material, kalyaan2023effect, pinilla2024survival}.}

\subsection{Comparison with Sellek et al. (2024)}

\AdrienRevision{During the revision of this manuscript, we became aware of \citet{sellek2024co2}, which also investigates the impact of pebble drift on molecular emission as seen with JWST. We briefly highlight here some similarities and differences. \citet{sellek2024co2} also uses 1D dust and volatile evolution models, with different scenarios for $v_\mathrm{frag}$, some of which lead to a traffic-jam in the inner disc. In addition to $\mathrm{H_2O}$, their models account for the evolution of $\mathrm{CO_2}$ ice and vapor and the presence of substructures, while the focus is on using the observable $\mathrm{CO_2}/\mathrm{H_2O}$ column density ratio. For water alone, and in agreement with our findings, \citet{sellek2024co2} also find that the presence of a traffic-jam leads to an almost constant $N_\mathrm{H_2O}^\mathrm{obs}(t)$, despite the inner disc being flooded in water (Scenario 1 in their Fig.~7). For this scenario, \citet{sellek2024co2} point out that using both $\mathrm{CO_2}$ and water could help break degeneracies, as the $\mathrm{CO_2}$ snowline is located further out and is unaffected by the presence of a traffic-jam at the water snowline. Additionally, they also generate synthetic spectra and perform slab model fits with a single temperature component (their Fig.~6), which at first glance appear not to pick up variations in cold water content (see their Appendix C.2). This may be partly due to the fact that \citet{sellek2024co2} assume a vertically isothermal temperature structure when converting their surface densities to column densities (Sect.~3.1), not capturing the (cold) gas-phase water reservoir directly above and outside the water snowline (see Fig.~\ref{fig:2D_dust_distribution}). Furthermore, they do not consider local chemical processing in the inner disc surface, captured in our approach by the $t_\mathrm{chem}$ parameter. In summary, where \citet{sellek2024co2} investigates the combined use of water and $\mathrm{CO_2}$ as a pebble drift tracer and our study focusses more on the radial and temperature gradients of water, both works highlight and bring into focus the complexities involved in connecting molecular emission lines from the disc surface to drifting solids in the disc midplane.}

\section{Conclusion}
\label{sec:conclusion}

In this work, we have studied how icy dust particles drifting from the outer disc lead to an enrichment of the inner disc in both solids and water vapor. We have developed a 1+1D approach to estimate the vertical and radial dust density distribution, based on dust models accounting for coagulation, fragmentation, and transport. From that, we derived the dust opacity and the height of the $\tau_\mathrm{20 \mu m} = 1$ surface, from which we derived the observable water reservoir and how it \AdrienRevision{evolves due to the interplay of radial transport (pebble delivery and advection/diffusion), vertical diffusion, and local processing in the exposed surface layers. Our findings are summarized as follows:}

\begin{enumerate}
    
    \item \AdrienRevision{The bulk reservoir of dust and water vapor present in the inner disc is deeply connected to the efficiency of pebble drift from the outer regions. High fragmentation velocity ($v_\mathrm{frag} \geq 10 \mathrm{~m~s^{-1}}$) favors the formation of large particles ($a_\mathrm{max} \geq 1 \mathrm{~cm}$, see Fig.~\ref{fig:1D_radial_dust_distribution}) that leads to a high but short-lived pebble flux through the disc (Fig.~\ref{fig:pebble_flux}).}

    \item \AdrienRevision{The height of the dust $\tau = 1$ surface is dependent two factors: (1) the dust density, that depends on the initial disc dust mass and the ability of the disc to maintain a large dust reservoir, e.g., due to inefficient radial drift or the presence of a traffic-jam effect in the inner regions, and (2) the dust vertical distribution, that is related to the dust scale-height, hence the dust particle size and disc turbulence. In the fragile dust model, the disc maintains a large optical depth for several Myr (see Fig.~\ref{fig:2D_dust_distribution}).}

    \item \AdrienRevision{Only a small fraction of the bulk water vapor may be observable above the dust $\tau = 1$ surface (Fig.~\ref{fig:f_obs}). The observable column density of water vapor varies with time and radial distance, and spans many orders of magnitude through the disc lifetime, where the actual value depends on the drift rate, the presence/absence of a traffic-jam, and the efficiency of UV and chemical processing (Fig.~\ref{fig:Fig4_column_density_vs_time} and Fig.~\ref{fig:Fig5_column_density_vs_radius}).}

    \item \AdrienRevision{The peak of observable vapor mass does not necessarily correlate with the actual peak in the bulk vapor reservoir (Fig.~\ref{fig:Fig4_column_density_vs_time}). In the case where a traffic-jam is present, the observable column density appears constant despite the intense delivery of vapor by large inward-drifting pebbles. Water vapor is effectively \textit{smuggled unnoticed}.}

    \item \AdrienRevision{Efficient radial drift and the absence of a traffic-jam uniquely leads in an observable water vapor mass that (Fig.~\ref{fig:FigY_obs_water_T_componant}): (1) may increase with decreasing disc dust mass \citep[][]{najita2013hcn, banzatti2020hints} and (2) has a linear trend with the current pebble flux through the snowline \citep[][Eq.~11]{romero2024retrieval}. However, in the presence of a traffic-jam, or when lines themselves become highly optically thick, a large range of pebble flux and disc dust mass may be associated to a similar observable reservoir.}
    
\end{enumerate}

\AdrienRevision{To make more quantitative comparisons with observed population-level trends \citep[e.g.,][]{banzatti2023jwst,banzatti2024atlas}, a larger parameter space exploration is required, notably exploring the impact of the disc set-ups (e.g., initial disc mass, disc radius, temperature structure, and turbulence level), and the impact of disc substructure and planetesimal formation on the pebble flux arriving at the water snowline \citep{kalyaan2023effect, easterwood2024water}. The results of such calculations can then be inspected with state-of-the-art retrieval methods capable of resolving temperature and column density gradients \citep[e.g.,][]{romero2024retrieval, kaeufer2024bayesian}. This will be the subject of future work.}

\section{Acknowledgment}

\AdrienRevision{We are grateful to the anonymous reviewer for their thorough and constructive comments that greatly improved the manuscript.} The results reported herein benefited from collaborations and/or information exchange within NASA’s Nexus for Exoplanet System Science (NExSS) research coordination network sponsored by NASA’s Science Mission Directorate and project ``Alien Earths'' funded under Agreement No. 80NSSC21K0593. A portion of this research was carried out at the Jet Propulsion Laboratory, California Institute of Technology, under a contract with the National Aeronautics and Space Administration (80NM0018D0004).

\section*{Data Availability}
The data generated by our dust evolution simulations based on \AdrienRevision{\texttt{chemcomp}} will be shared upon reasonable request.

\appendix 

\section{Optical depth and $\tau = 1$ surface}
\label{sec:appendix_opacity}

To compute the dust optical depth, we need to estimate the total opacity of the dust size distribution at each point of the radial and vertical structure. To do so, we use the \texttt{DSHARP-OPAC} package developed by \citet{birnstiel2018disk}. We assume dust particles to be made of a mixture of silicates, troilite, and refractory organics with a relative mass abundance respectively of $0.411$, $0.093$, and $0.496$. Their internal densities are taken to be $3.3 \mathrm{~g~cm^{-3}}$, $4.83 \mathrm{~g~cm^{-3}}$, and $1.50 \mathrm{~g~cm^{-3}}$, respectively, leading to a bulk density $\rho_\mathrm{s} = 2.11 \mathrm{~g~cm^{-3}}$. Outside of the water snowline, dust grains are additionally covered in water ice. \AdrienRevision{Except right outside the snowline, where vapor re-condensation enhances the ice fraction on grains, the mass fraction of water ice is typically equal to the initial water-to-refractory ratio $\delta_\mathrm{w2r, 0}=0.2$. For simplicity, we use this value when computing opacities outside the midplane snowline ($r > r_\mathrm{snow}^\mathrm{mid}$)}. As the internal density of water ice is $0.92 \mathrm{~g~cm^{-3}}$, the bulk density of icy grains outside of the water snowline is $\rho_\mathrm{s}=1.67 \mathrm{~g~cm^{-3}}$. Following that dust composition, the optical constants are taken from \citet{warren2008optical} for water ice, \citet{draine2003interstellar} for astronomical silicates, and \citet{henning1996dust} for troilite and refractory organics. With \texttt{DSHARP-OPAC}, we compute the optical constants of the dust mixture using the Bruggeman effective theory, and then calculate the size-dependent absorption and scattering opacity using Mie theory. Note that we are using the effective anisotropic scattering $\kappa_\nu^\mathrm{sca, eff} = \kappa_\nu^\mathrm{sca} (1-g_\nu)$, where $g_\nu$ is the forward scattering adjustment factor. With \texttt{chemcomp} and our assumption of vertical equilibrium between settling and turbulent mixing, we are resolving at all time the dust size distribution at each point of our radial and vertical grid. We use these results to compute the total opacity of the dust distribution $\kappa_\nu^\mathrm{tot}(r, z, t)$, which is then used in Equation~\ref{eq:optical_depth} to estimate the optical depth due to the dust continuum.

\bibliographystyle{mnras}
\bibliography{paper.bib} 

\bsp	
\label{lastpage}
\end{document}